\documentclass[%
prx,
reprint,
amsmath,amssymb,
nofootinbib,
aps,
]{revtex4-2}

\usepackage{times}
\usepackage{graphicx}
\usepackage{dcolumn}
\usepackage{bm}
\usepackage{amsmath,amssymb}
\usepackage{physics}
\usepackage[normalem]{ulem}
\usepackage{url,hyperref}
\hypersetup{colorlinks,linkcolor=black,citecolor=airforceblue,urlcolor=airforceblue,breaklinks=true}

\usepackage{fixme}
\usepackage{extarrows}
\usepackage{xcolor}
 \fxsetup{status=draft}
 \fxsetup{inline=false, margin=true}
 \fxsetup{theme=color, author=GF}
\DeclareGraphicsExtensions{.png,.jpg,.eps,.pdf}

\makeatletter
\newcommand{\fmarki}{*}
\newcommand{\fmarkii}{\ensuremath{\dagger}}
\def\@fnsymbol#1{{\ifcase#1\or \fmarki\or \fmarkii \else\@ctrerr\fi}}
\makeatother

\renewcommand{\fmarki}{$\dagger$}
\renewcommand{\fmarkii}{$\ddagger$}

\begin{document}
\definecolor{airforceblue}{rgb}{0.36, 0.54, 0.66}

\definecolor{darkblue}{rgb}{0.0, 0.0, 0.55}
\definecolor{grey}{rgb}{0.5, 0.5, 0.5}  % A medium grey

\title{Fast gates for bit-flip protected superconducting qubits }

\author{C. A. Siegele$^\star$}
\email{christian.siegele@ist.ac.at}
\author{A. A. Sokolova$^\star$}
%\email{alesya.sokolova@ist.ac.at}
\author{L. N. Kapoor}
\author{F. Hassani}
\author{J. M. Fink}
\email{jfink@ist.ac.at}
\affiliation{Institute of Science and Technology Austria, 3400 Klosterneuburg, Austria}

\date{\today}

\begin{abstract}  

Superconducting qubits offer an unprecedentedly high degree of flexibility in terms of circuit encoding and parameter choices. However, in designing the qubit parameters one typically faces the conflicting goals of long coherence times and simple control capabilities. Both are determined by the wavefunction overlap of the qubit basis states and the corresponding matrix elements. Here, we address this problem by introducing a qubit architecture with real-time tunable bit-flip protection. In the first,  the `heavy' regime, the  energy relaxation time can be on the order of hours  for fluxons located in two near-degenerate ground states, as recently demonstrated in  Ref.~[Hassani et al., Nat.~Commun.~14 (2023)]. The second,  `light' regime, on the other hand facilitates high-fidelity control on nanosecond timescales without the need for microwave signals. We propose two different tuning mechanisms of the qubit potential and show that base-band flux-pulses of around 10 ns are sufficient to realize a universal set of high-fidelity single- and two-qubit gates. We expect that the concept of real-time wavefunction control can also be applied to other hardware-protected qubit designs. 

\end{abstract}

\maketitle
%\section{Introduction}
\def\thefootnote{$\star$}
\footnotetext{These authors contributed equally to this work.}
\def\thefootnote{\arabic{footnote}}

\section{Introduction}

Superconducting qubits have significantly contributed towards the development of a universal quantum computer \cite{review_general}. The most prominent one  is the transmon, given the simple fabrication and protection against charge noise-induced dephasing \cite{koch_transmon}. Despite the significant results that have been demonstrated with transmon qubits, the low anharmonicity, as well as relaxation due to dieletric loss and material defects, limit their performance. Therefore, different types of superconducting  qubits such as the fluxonium qubit - an rf-SQUID type circuit with enhanced anharmonicity - have been  developed \cite{first_fluxonium}. Recently, high-fidelity
 single-qubit and two-qubit gates have been investigated theoretically \cite{schuster_koch_theory_2q, Rosenfeld,vavilov, vavilov2, ustinov} and implemented experimentally \cite{manucharyan_2q,deng_2q, wang_2q,gral,schuster_1q,schuster_flux,high_coherent,oliver_two_fluxonium_transmon, simakov} and the
 potential of fluxoniums as a scalable architecture  demonstrated
\cite{fluxonium_processor}. Fluxonium qubits have higher characteristic impedance compared to flux qubits and can be realized in two main regimes: coined  `heavy'
 and `light' \cite{moving_beyond_transmon, Peruzzo2021}. In the heavy regime, transitions between the two lowest fluxon states are exponentially suppressed with the energy barrier separating them. This intrinsic  protection however complicates active control of the qubit. In the light regime on the other hand, the matrix elements are substantially larger and the reduced energy dispersion leads to a smaller flux-noise sensitivity. This regime - with the Blochnium being the most extreme   - is more suitable for control at the cost of lower bit-flip protection \cite{blochnium,Peruzzo2021}. A special case of an `ultra-heavy' fluxonium, is the inductively-shunted  transmon (IST) qubit - an intermediate-scale impedance rf-SQUID with a large capacitance and a large inductance. In this case, a relaxation time between the two fluxon states in the range of hours has been observed \cite{farid_fluxon} close to the half-flux point. However, due to the  vanishingly small matrix elements, coherent control of the qubit subspace has not been demonstrated experimentally. \\

In this work, we propose to realize gates between the exponentially  bit-flip-protected fluxon states by real-time tuning the degree of protection: between the heavy regime with long  $T_1$ times for information storage and the light regime that allows fast high-fidelity  gates due to the increased matrix elements. Moreover, we show that the mere act of tuning the qubit into the light regime for an optimized duration, implements a single- or two-qubit gate of choice, up to local phase rotations. With this approach, the advantages of both regimes are combined in a single circuit. Experimentally, fast flux control has already been used for the implementation of high-fidelity single qubit gates \cite{schuster_1q}, by the means of non-adiabatic Landau-Zener transitions \cite{oliver_labdau_zenner}. In contrast to ramping the external flux bias, we propose an in-situ tuning of the Josephson energy  \cite{Poletto2009,Poletto2009b} or alternatively the effective mass. This approach is conceptually related to the single qubit gates proposed for the oscillator-stabilized flux qubit \cite{koch2006, koch2005}, as well as the one proposed in \cite{flux_qubit_switch} for flux qubits.

\begin{figure}[h!]
\centering\includegraphics[width=\linewidth]{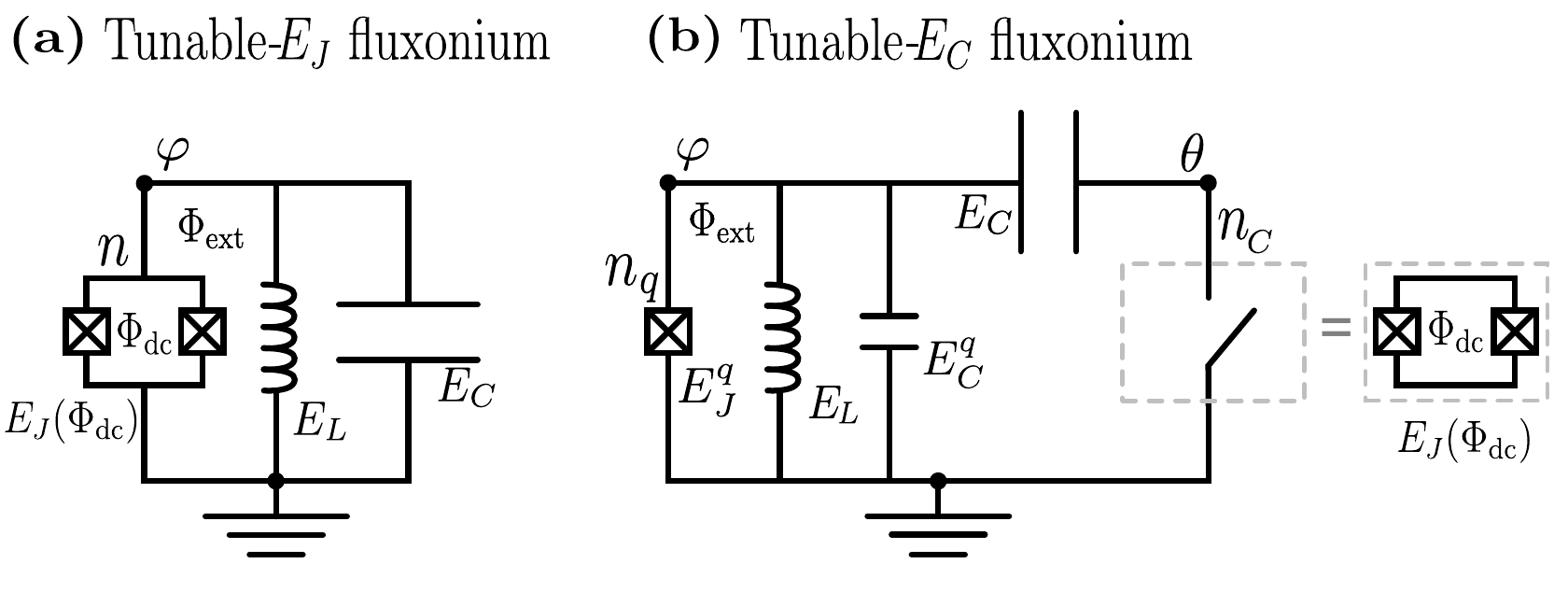}
\caption{\textbf{Tunable circuits.} {(a)}, Tunable-$E_J$ fluxonium: The Josephson energy $E_J$ of the dc-SQUID  is tuned through the flux $\Phi_\textrm{dc}$. $E_L$ being  the inductive energy, $E_C$ the capacitive energy, $n$ the excess charge on the island, $\varphi$  the phase variable and $\Phi_\textrm{ext}$ is the flux in the rf-SQUID. {(b)}, Tunable-$E_C$ fluxonium: $E_J^q$ represents the Josephson energy, $E_L$ is the inductive energy, and $E_C^q$ is the capacitive energy of the qubit in the light regime. $E_C$ is the capacitive energy of an additional large shunt capacitance connected to a dc-SQUID with a tunable $E_J$ value, acting as a 'switch'. $n_q$ is the excess charge on the rf-SQUID island, $n_c$ the charge on the dc-SQUID island, and $\varphi$ and $\theta$ are the corresponding phases in the two loops.}\label{scheme}
\end{figure}

\section{Tunable circuits }

%\textcolor{darkblue}{

We propose two different circuits that allow fast gate operations between the two fluxon states with disjoint support in the $\varphi$-coordinate. In \autoref{scheme}(a), the first circuit depicts the {tunable-}$E_J$ {fluxonium}, for which the potential landscape of the qubit is tuned in-situ. Compared to an ordinary fluxonium, the single Josephson Junction (JJ)  is replaced by a dc-SQUID, allowing to alter the effective value of the Josephson energy $E_J$, by  tuning its external magnetic flux  $\Phi_\textrm{dc}$. The 'heavy' regime ($E_J/E_C\gg 1$) is attained  for a high value of  $E_J$, i.e.~the external magnetic flux in the dc-SQUID being $\Phi_\textrm{dc}/\Phi_0=0$, where $\Phi_0$ is the magnetic flux quantum. Oppositely for $\Phi_\textrm{dc}/\Phi_0\approx0.5$, the circuit is in the 'light' regime ($E_J/E_C \sim 10$). The circuit Hamiltonian can be written as

\begin{equation}\label{H_EJn}
    \hat{H}_{E_J} =E_C (2 \hat{n})^2 + \frac{E_L}{2} (\hat{\varphi}- \varphi_\textrm{ext})^2 -E_J (\varphi_\textrm{dc}) \cos (\hat{\varphi}),
\end{equation} 

with $\varphi_\textrm{dc} = 2\pi \Phi_\textrm{dc}/\Phi_0$ being the flux inside the dc-SQUID and $E_J(\varphi_\textrm{dc}) = E_{J, \textrm{max}}|{\cos(\frac{\varphi_\textrm{dc}}{2})}|$, thus forming two first-order insensitive working points at zero and half flux. For clarity, a possible asymmetry of the Josephson junctions in the dc-SQUID is omitted here. In the presence of a finite asymmetry, the Hamiltonian can be re-written in a similar form, up to an additional $\Phi_\textrm{dc}$-dependent  correction to the reduced external flux $\varphi_\textrm{ext}=2\pi \Phi_\textrm{ext}/\Phi_0$,
 that can be adjusted for, as shown in the full derivation in Appendix~\ref{app}. In this case,  $E_J/h$ can not fully be tuned to zero, but a finite value, which we account for by considering a range of $1$-$12$ GHz.\\

\begin{figure}[t]
\vspace{0.5cm}
\centering\includegraphics[width=\linewidth]{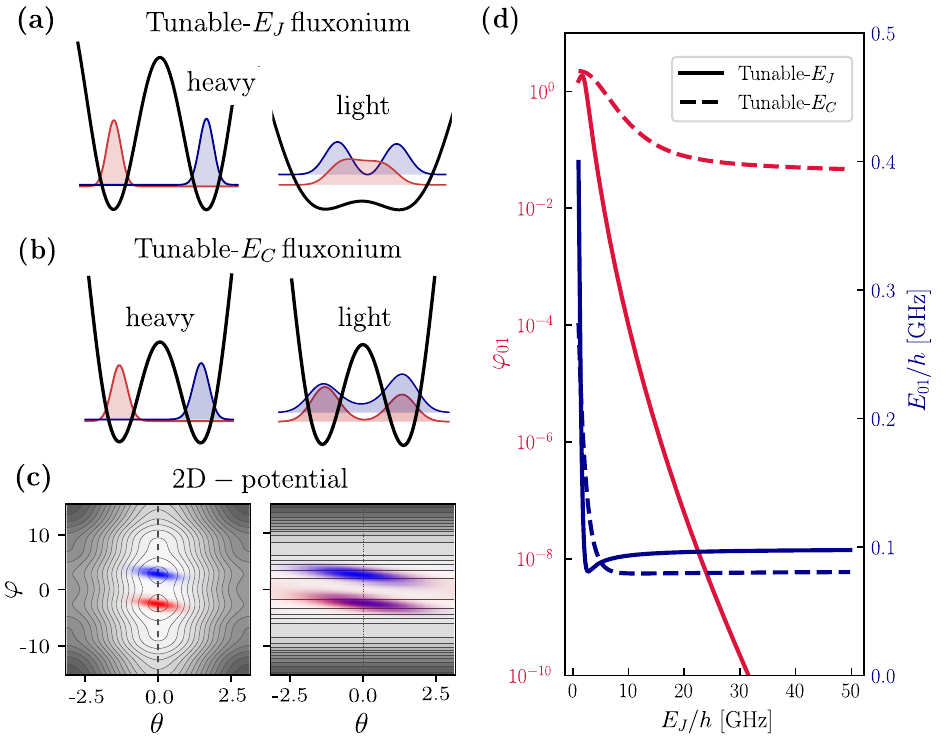}
\caption{\textbf{Qubit properties at flux working point}  $\Phi_\textrm{ext}/\Phi_0 = 0.495$.
{(a)} %1-D 
Potential and absolute value of  wavefunctions of the first two fluxon levels (red and blue) of the tunable-$E_J$ fluxonium with $E_C/h = 0.25$~GHz, $E_L/h = 0.5$~GHz, $E_J/h = 1$~GHz in the light regime and 12~GHz in the heavy regime. 
{(b)} $\theta=0$ cross-section of the 2d-potential and wavefunctions of the tunable-$E_C$ fluxonium with $E_C/h = 0.15$~GHz, $E_C^q/h = 1$~GHz, $E_L/h = 0.5$~GHz, $E_J^q/h = 3$~GHz and $E_J/h$ of the dc-SQUID ranging from 1 to 50~GHz. {(c)} 2D-potential (gray scale and contour lines) and wavefunctions of the tunable-$E_C$ fluxonium. Dashed lines indicate the cross-section $\theta=0$ shown in (b). 
{(d)} Phase matrix elements $\varphi_{01}$ (red) and transition frequencies $E_{01}/h$ (blue) between the two lowest fluxon states $\ket{0}$ and $\ket{1}$ for the tunable-$E_J$ (solid) and tunable-$E_C$ (dashed) circuit as a function of $E_J$ of the dc-SQUIDs.} 
\label{wavefunctions_fig}
\end{figure}

In the case of the {tunable-}$E_C$ {fluxonium}, shown in \autoref{scheme}(b), the $\varphi$-dependent part of the potential is unperturbed and instead  the effective mass is tuned to increase the wavefunction overlap of the fluxon states.  Since tuning a capacitance itself on a nanosecond scale is difficult, we propose to effectively connect and disconnect a large capacitance to the fluxonium circuit. This `switch' is similarly realized through a dc-SQUID, being equivalent to a short for a low inductance around $\Phi_\textrm{dc}/\Phi_0=0 $ and equivalent to an open for $\Phi_\textrm{dc}/\Phi_0=0.5 $. The Hamiltonian for this circuit can be written as 

\begin{align}\label{H_EC}
        \hat{H}_{E_C} = &E_C^q (2 \hat{n}_q)^2+ \frac{E_L}{2} (\hat{\varphi}- \varphi_\textrm{ext})^2 -E_J^q \cos \hat{\varphi}  \\\nonumber &+ 4 (E_C^q + E_C) (\hat{n}_C - n_g)^2- E_J(\varphi_\textrm{dc}) \cos \hat{\theta}  \\\nonumber &-8 E_C^q \hat{n}_C \hat{n}_q.
\end{align}
 In addition to the fluxonium phase $\hat{\varphi}$, another phase over the dc-SQUID $\hat{\theta}$ appears. This gives a two-dimensional potential as displayed in \autoref{wavefunctions_fig}(c), with the cross-section  over ${\varphi}$ remaining an ordinary fluxonium potential, but in the $\theta$ coordinate  the wavefunctions are delocalized  in the light regime. From \autoref{H_EC}, this circuit can be  equivalently interpreted as a fluxonium capacitively coupled to a transmon with $E_C^t = E_C^q+E_C$, with $\hat{\theta}$ representing the phase of this effective transmon mode.\\

Figure~\ref{wavefunctions_fig}(d) shows the transition energy $E_{01}$ and phase matrix element $\varphi_{01} = |\bra{0} \hat{\varphi} \ket{1}|$ between the two lowest energy fluxon states $\ket{0}$ and $\ket{1}$ as a function of the tuning parameter $E_J(\varphi_\textrm{dc})$ of the dc-SQUID for both proposed qubits. The phase matrix element $\varphi_{01}$ is proportional to the charge matrix element $\bra{0} \hat{\varphi} \ket{1} = \frac{8E_C}{\omega}\bra{0} \hat{n} \ket{1}$ \cite{schuster_flux} and  an  indicator for the degree of protection from energy relaxation in the fluxon basis. The energy splitting of the qubit states $E_{01}$ is quite similar for both qubits as a function of $E_J(\varphi_\textrm{dc})$, however  $\varphi_{01}$ is substantially stronger suppressed for the tunable-$E_J$ fluxonium.

\section{Single-qubit gates}\label{section_1q}

Typically, superconducting qubits are controlled by microwave pulses, whereas in this proposal, all required gates can be performed solely with base-band flux pulses without IQ-modulation. Physically, the qubits are tuned to the light regime by applying a flux pulse to the dc-SQUID that acts as a switch for the tunable-$E_C$ fluxonium and lowers the potential barrier for the tunable-$E_J$ fluxonium.\\

\begin{figure}[t]
\centering\includegraphics[width=\linewidth]{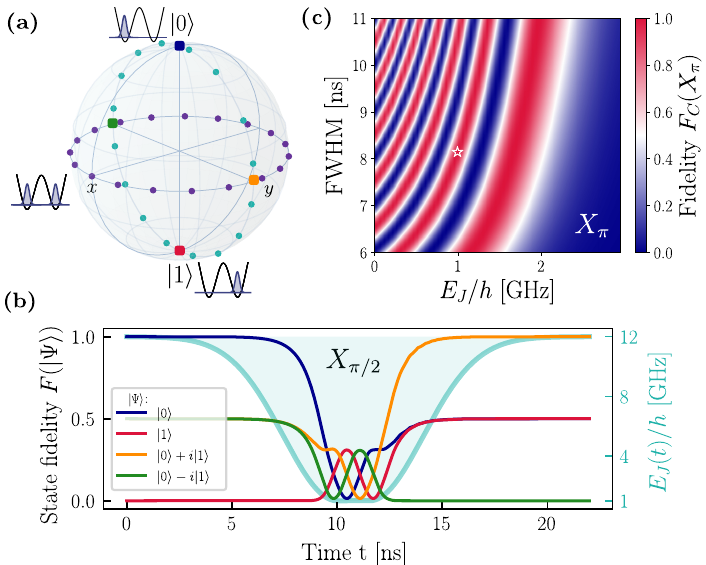}
\caption{
\textbf{Flux-pulse control of the tunable-$E_J$ fluxonium}.
{(a)}, Bloch sphere representation of the qubit subspace with $\ket{0},\ket{1}$ being the protected fluxon states. The purple points depict a  rotation  around $\sigma_z$-axis, performed via detuning from the external  flux  point $\Phi_\textrm{ext}/\Phi_0=0.50$. Turquoise points indicate a rotation around the $\sigma_x$-axis, performed through the flux pulse of flat-top length $l_\textrm{flat}$. {(b)}, State Fidelity $F(\ket{\Psi})$ visualized over time of the initial state $\ket{0}$ throughout  a $X_{\pi/2}$ gate. The optimized, flat-top Gaussian flux pulse (turquoise curve) that varies  $E_J/h$ from $12$~GHz to $1$~GHz, is indicated by the secondary axis. Colored cardinal points on the Bloch sphere correspond to the individual labeling of the curves.
{(c)}, Closed-system  gate fidelity $F_C(U_T=X_{\pi})$  as a function of the FWHM  and  $E_J/h$, with  $l_\textrm{flat}$ of the pulse being varied between $0-5$~ns. The white asterisk indicates the an optimized parameter set.}
\label{single_qubit_fig}
\end{figure}

At the flux point $\Phi_\textrm{ext}/\Phi_0=0.5$, the potential of the {tunable-}$E_J$ {fluxonium} is well approximated by a double-well potential and the lowest eigenstates are the symmetric and anti-symmetric superpositions  of the fluxon states localized in individual wells. We use the convention for which those are identified as  $\ket{+},\ket{-}$  and the computational basis $\ket{0},\ket{1}$  as the left and right fluxon states, as visualized on the Bloch sphere in \autoref{single_qubit_fig}(a) (up to an exponentially small correction, as for cat states \cite{mirrahimi}). By modifying the potential landscape in a diabatic manner, the qubit states in the heavy-regime $\{e^h_i\}$ are no longer eigenstates of the  Hamiltonian in the light regime $\{e^l_i\}$ and start to precess. In this choice of basis, a pulse lowering the potential barrier realizes a continuous rotation around the $\sigma_x$-axis of the Bloch sphere  with respect to the eigenstates of the idle Hamiltonian. Turquoise points on the Bloch sphere represent respective final states after a   flat-top Gaussian flux-pulse (as depicted in \autoref{single_qubit_fig}(b)) of varied 
flat-top length $l_\textrm{flat}$, 

which sets  the rotation angle. For universal qubit control, an additional, continuous rotation around the $\sigma_z$-axis can be performed, by tuning $\Phi_\textrm{ext}$ away from the half flux point for a finite duration, which introduces an asymmetry to the potential that lifts the energy splitting of the fluxon states temporarily. This rotation on the equator is depicted by purple points for different durations.\\

In the numerical simulations, the following realistic parameters were used $E_C/h = 0.25$~GHz, $E_L/h = 0.5$~GHz, and $E_J/h$ ranging from 1 to 12~GHz. Therefore, in the idle mode the circuit is situated in-between a heavy fluxonium \cite{schuster_1q} and the IST-qubit \cite{farid_fluxon} limit, and for the short duration of the single- and two-qubit gates between a typical fluxonium and a flux qubit. The flat-top Gaussian pulse is characterized by its full width at half maximum (FWHM), with a  flat-top length $l_\textrm{flat}$ and an amplitude in terms of $E_J/h$, as shown in  \autoref{single_qubit_fig}(b) for an optimized $X_{\pi/2}$ gate. We extract the gate fidelity by reconstruction of the Pauli process matrix $\chi_R$ \cite{tomography_intro} with respect to the one for the ideal gate  $\chi$ using

\begin{equation}\label{fidelity_pauli_eq}
    F = \frac{2\Tr (\chi^T \chi_R)  + 1}{3}.
\end{equation}

In the closed-system case, the infidelity results in $1-F_c=3 \times 10^{-6}$ for a  $X_{\pi}$ gate with the optimal parameters being $\mathrm{FWHM}=8$~ns, $l_\textrm{flat}=2$~ns and $E_J/h=0.963 $~GHz. The Hilbert space of the fluxonium in the simulation of the tunable-$E_J$ qubit is truncated to 100 Fock states. In the open-system case, the time evolution is simulated in the presence of noise, that is modeled using the expected decoherence rates presented in \autoref{sec_coherence}, and the system is truncated to the qubit subspace. For the specified optimal parameters, we find $1-F_O =2 \times 10^{-3}$, with the main infidelity contribution due to dephasing in the beginning and end of the pulse. Figure~\ref{single_qubit_fig}(c) depicts the $X_{\pi}$-gate fidelity $F_C$ as a function of the FWHM  and $E_J$ with the optimized gate parameters indicated by an asterisk. Multiple maxima highlight that for a given pulse length an optimal pulse amplitude in terms of $E_J$  can be found to maximize the gate fidelity. \\

In the heavy regime, we expect a significant reduction of $T_1$ times approaching the half-flux bias point 
due to hybridization of higher excited states that dominate bit flip times at finite temperature in the heavy regime, as discussed in section 
\autoref{sec_coherence}. For this reason, here we demonstrate that the proposed flux-gates can also be performed away from the half-flux point for optimized flux pulse parameters. If we do not introduce a new rotating frame, the Bloch vector rotates around the $\sigma_z$-axis in the idle-mode, due to the finite energy level splitting between the fluxon states. Choosing the working point  
$\Phi_\textrm{ext}/\Phi_0=0.495$, we simulate the Hadamard gate ($\pi$ rotation around the axis $\sigma_H=(\sigma_x+\sigma_z)/\sqrt{2}$) with a flat-top Gaussian pulse (similar to \autoref{single_qubit_fig}(b)) with $\mathrm{FWHM}=7.5$~ns, $l_\textrm{flat}=1.5$~ns and $E_J/h=0.9872 $~GHz, 
resulting in the closed-system gate-infidelity of $1-F_C=2 \times 10^{-5}$. \\

\begin{figure*}[t]
\centering\includegraphics[width=\linewidth]{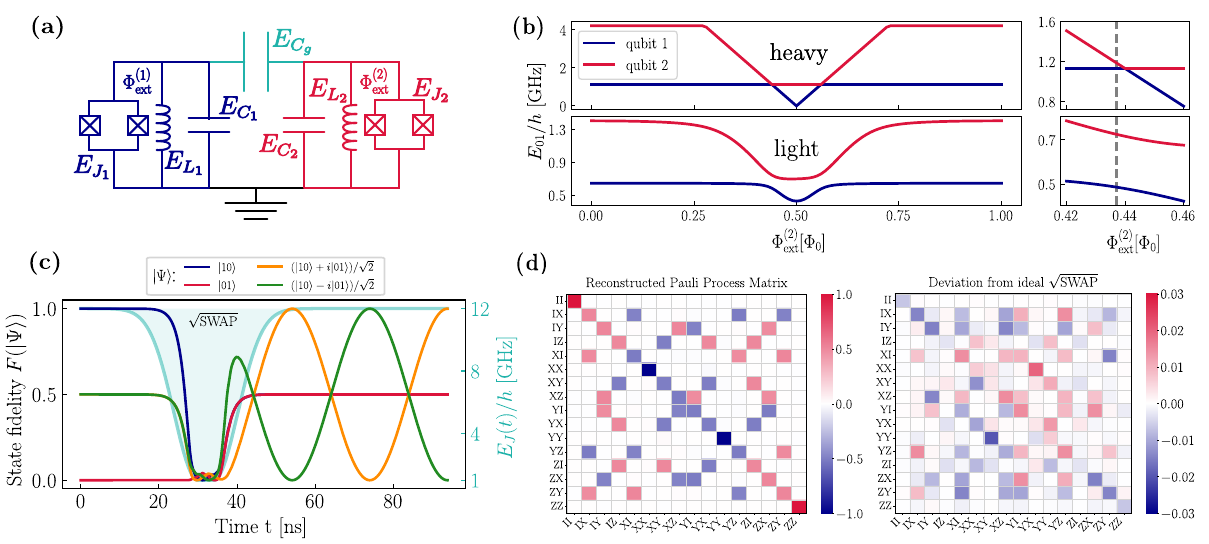}
\caption{\textbf{Two-qubit gate scheme} {(a)}, Electrical circuit diagram for two capacitively-coupled tunable-$E_J$ qubits. {(b)}, Energy spectrum for the qubits operated in the heavy (top), and light (bottom) regime. The abscissa represents the external flux of the second qubit $\Phi_{\mathrm{ext}}^{(2)}$. 
 with $\Phi_{\mathrm{ext}}^{(1)} = 0.44~ \Phi_0$ being fixed. In the zoomed-in panels  in the right column, the grey dashed line depicts the working point position for the $\sqrt{\text{SWAP}}$ gate around the anti-crossing. {(c)},  Fidelity of the state evolution throughout the flux-pulse of the initial state $\ket{10}$. Oscillations in the \{$\ket{10},\ket{01}$\} manifold are due to the energy splitting between the two states in this frame.   The flat-top Gaussian pulse shape applied to both qubits  is displayed in turquoise in terms of $E_J/h$. {(d)}, Left: Reconstructed Pauli process matrix for the simulated optimized $\sqrt{\text{SWAP}}$ gate. Right: Deviation with respect to the ideal target  $\sqrt{\text{SWAP}}$ gate.}\label{2_qubit_fig}
\end{figure*}

A similar optimization can be performed for the tunable-$E_C$ fluxonium, for which we find a closed system infidelity of $1-F_C = 1 \times 10^{-3}$ for the Hadamard gate at the same bias flux working point. This result was obtained with more coarse optimization routine due to the substantially larger Hilbert space involved. While we believe that higher fidelities could be achieved in principle, 
we expect the achievable optimum to be lower compared to the tunable-$E_J$ qubit, in part due to less well localized basis states in the light regime, for the experimental parameters considered. Moreover, the open system fidelity is also expected to suffer from charge noise due to the additional qubit island in this circuit. In general, the introduced method of tuning the wavefunction overlap can in principle always be complemented by additional microwave pulses in the light regime to further push the gate fidelity towards unity. This would be at the expense of additional required RF-control capabilities.

\section{Two-qubit gates}

For the implementation of a two-qubit gate, a simple capacitive coupling between the two qubits suffices, since the interaction is tunable through the individual dc-SQUIDs, as visualized in \autoref{2_qubit_fig}(a) for the tunable-$E_J$ fluxonium. The coupling capacitance is chosen, such that the two qubit sub-spaces can be considered decoupled when they are operated in the heavy regime  due to the vanishingly small respective charge matrix element. Similarly as for the single qubit-gate, tuning both qubits to the light regime leads to a hybridization of the qubit states and allows fast, high-fidelity two-qubit gates. For both qubits, the energy values are chosen close to the single qubit case, but slightly distinct. The parameters of the first qubit are $E_{C_1}/h = 0.30$ GHz, $E_{L_1}/h = 0.52$ GHz and of the second $E_{C_2}/h = 0.30$ GHz, $E_{L_2}/h = 0.50$ GHz. The tunability of the respective $E_{J_i}/h$ value  is considered in the range of $1-12$ GHz, identical to  the single-qubit case. The joint Hamiltonian of the capacitively coupled qubits is given by 

\begin{equation}
    \hat{H}_{12} =\hat{H}_{1}+\hat{H}_{2}+8 \frac{E_{C_1}E_{C_2}}{E_{C_1}+E_{C_2}+E_{C_g}}\hat{n}_1 \hat{n}_2, 
\end{equation}

with $\hat{H}_{i}$  being the respective single qubit Hamiltonian in ~\autoref{H_EJn} and $E_{C_g}/h = 6.67$ GHz is the energy of the coupling capacitor. \\

 In \autoref{2_qubit_fig}(b), the energy spectrum of the two capacitively coupled qubits in the heavy and light regime is displayed versus the external flux  $\Phi_\mathrm{ext}^{(2)}$ of qubit 2, while for  qubit 1 being fixed to $\Phi_\mathrm{ext}^{(1)} = 0.44 ~\Phi_0$. By tuning both qubits to the light regime the anti-crossing is around  239 MHz for $\Phi_\mathrm{ext}^{(2)} = 0.437 ~\Phi_0$.  In principle, any value close to $\Phi_\mathrm{ext}^{(1)} = 0.5 ~\Phi_0$ is a valid working point, since the size of the anti-crossing remains of the order of $250-400$ MHz at any flux point. At this working point, a flat-top Gaussian pulse is applied to the individual dc-SQUID loops, as depicted in \autoref{2_qubit_fig}(c). The flux-pulse is  optimized with regard to the $\sqrt{\text{SWAP}}$ gate, that forms, together with individual qubit control, a universal quantum gate set. The optimal parameters found for the $\sqrt{\text{SWAP}}$ gate are: a $\mathrm{FWHM} = 19$ ns of the pulse with a flat-top part of $l_{\mathrm{flat}}=4$~ns, the external fluxes of both qubits set to  $\Phi_\mathrm{ext}^{(1)} = 0.4399~ \Phi_0$, $\Phi_\mathrm{ext}^{(2)} = 0.4369 ~\Phi_0$ and the minimum value of $E_{J_i}/h=1.0068$~GHz. For ease of optimization we kept the pulse parameters of both qubits the same, but individual fine-tuning might improve the gate fidelity even further. We reconstruct the full Pauli process matrix of the $\sqrt{\text{SWAP}}$ gate in the closed system case, as visualized in \autoref{2_qubit_fig}(d) and extract an infidelity of $1-F_C = 3 \times 10^{-4}$. In the numerical simulations, the Hilbert space truncation is set to 50 Fock states. In the open-system case, we find an infidelity $1-F_O = 6 \times 10^{-3}$, which is limited by the short dephasing times as in the single-qubit case. The $\sqrt{\text{SWAP}}$ gate was the choice for the two-qubit gate as it resulted in the highest fidelity, however other two-qubit gates, i.e. ${\text{iSWAP}}$ or $\sqrt{\text{iSWAP}}$ can be realized by adjusting the flux-pulse parameters. The two-qubit gate simulations for the tunable-$E_C$ qubit are left for future work. Notably, this proposed in-situ tuning of the Josephson energy gives the advantage of realizing high-fidelity two-qubit gates on the nano-second scale without the need of additional circuit elements as in other architectures, i.e. in the form of tunable couplers \cite{schuster_koch_theory_2q, oliver_two_fluxonium_transmon}.

\section{Expected coherence properties}\label{sec_coherence} 

In this section, we present the coherence properties for the  {tunable-}$E_J$ {fluxonium}. Since the parameters of the proposed qubits, are close to the one of the IST-qubit in \cite{farid_fluxon}, similarly long relaxation times of the fluxon states are expected, however not in the range of hours due to the smaller value of $E_J$. Typically for fluxonium devices, the coherence times are estimated using  Fermi's Golden rule, by computation of the matrix element of a respective noise operator $\bra{0} \hat{O} \ket{1}$ and the associated noise spectral density $S(\omega)$. However, given the very small matrix element of the fluxon states in the heavy regime, direct transitions between the two qubit states are estimated to occur on timescales of years, but do not represent the dominant decay channel. % at finite temperature. 
Instead, leakage to higher levels within a well, e.g. due to finite temperature, results in incoherent multi-level decay processes. Here, we consider the full energy level structure of the fluxonium Hamiltonian and compute the transition rates between the respective levels. Given the experimental results for fluxonium devices in \cite{Sun2023,mencia2024}, we estimate the $T_1$ times to be limited by dieletric loss. The transition rate between two respective levels  ${\Gamma}^{\mathrm{diel}}_{ij}$ for dielectric loss is computed using

\begin{align}
    {\Gamma}^{\mathrm{diel}}_{ij} = \tfrac{\hbar \omega_{ij}^2}{4 E_CQ_{\mathrm{cap}}( \omega_{ij}) } |\bra{i} \hat{\varphi} \ket{j}|^2 \frac{\text{coth}(\tfrac{\hbar |\omega_{ij}|}{2k_\mathrm{B} T_{\mathrm{eff}}})}{1+\text{exp}(-\tfrac{\hbar \omega_{ij}}{k_\mathrm{B} T_{\mathrm{eff}}})},
\end{align}

where the capacitive quality factor is assumed to be $Q_{\mathrm{cap}}( \omega_{ij})=10^5 \times (\frac{2\pi \times 6~\mathrm{GHz}}{\omega_{ij}})^{0.7}$ and an effective bath temperature $T_{\mathrm{eff}}$ of 60 mK and 70 mK respectively. For the estimates of $T_1$, we simulate the time evolution in the eigenstate basis of the fluxonium Hamiltonian under the Lindblad equation with jump operators being defined as $L_{ij}=\sqrt{\Gamma_{ij}} \ket{j}\bra{i}$,  to include all possible decay channels and extract the value of $T_1$ from the 
 main population decay. Due to the multiple decay channels, the population does not undergo a single exponential decay, however other quantities can be used for decoding {\cite{ruiz, fluxcat}, that allows extracting a single decay constant.   \\ 
 
\begin{figure}[t]
\centering\includegraphics[width=\linewidth]{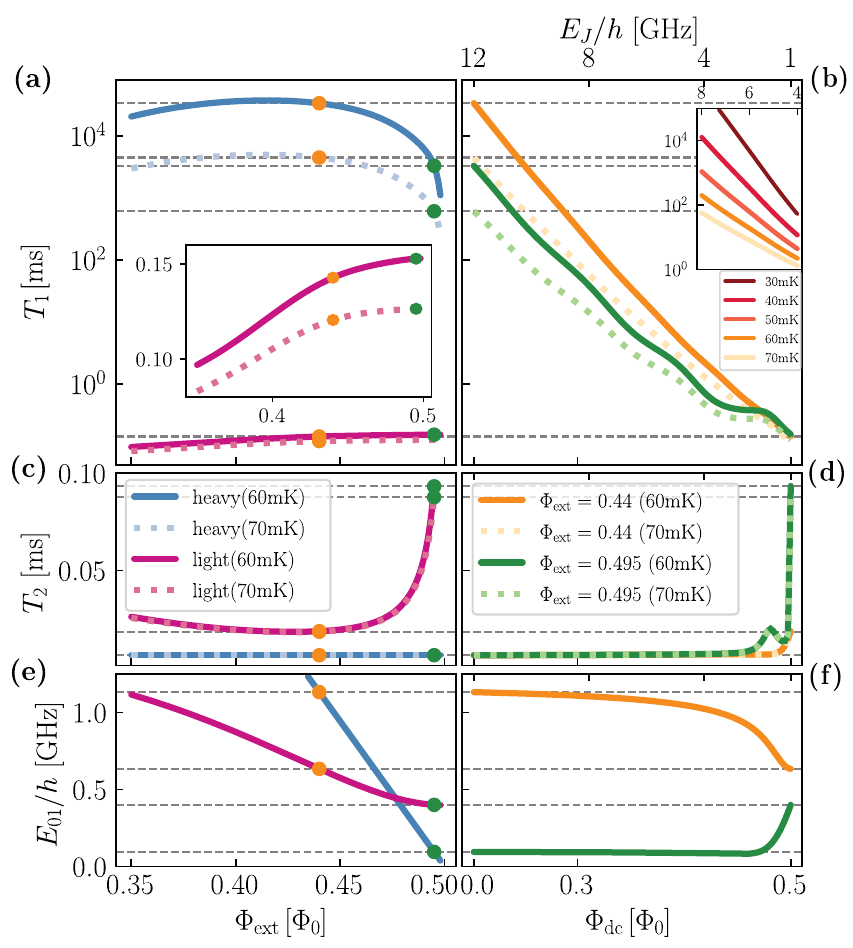}
\caption{\textbf{Coherence properties  for the  tunable-$E_J$ fluxonium.} Left column: Coherence time values $T_1$ in {(a)} and $T_2$  in {(c)} as well as the transition frequency $E_{01}/h$ in (e) as a function of $\Phi_\mathrm{ext}$ in the heavy (blue) and light (purple) regime for two different temperatures $60$ mK and $70$ mK. The disc-symbols indicate the working points of the single qubit gate (green) $\Phi_{\mathrm{ext}} = 0.495 \, \Phi_0$ and the two-qubit gate $\Phi_{\mathrm{ext}} = 0.44 \,\Phi_0$ (orange). Right column: Coherence times $T_1$ in {(b)} and $T_2$  in {(d)} as well as the transition frequency $E_{01}/h$ in {(f)} versus $\Phi_\mathrm{dc}$ that tunes the effective value of $E_J/h$ (top axis) in the range of $1-12$ GHz. Curves for the working points of the single and two-qubit gates are displayed in green and orange respectively. The inset in panel {(b)} shows $T_1 $ versus $E_J/h$ at $\Phi_{\mathrm{ext}} = 0.44 \,\Phi_0$ for additional temperatures. The dashed grey lines indicate the vertical position of the working points.}\label{coherence_fig}
\end{figure}

The values of $T_1$ are visualized in \autoref{coherence_fig}(a) as a function of  the external flux $\Phi_\mathrm{ext}$   for the two temperatures in the heavy and light regime. Around the working points of the single- and two-qubit gates (indicated by symbols) the immense difference between the two regimes  by a factor of around $~10^5-10^6$ is visible. In the heavy regime, the expected $T_1$ drops by more than an order of magnitude  approaching the half-flux point, which is attributed to increased  hybridization of higher-lying states with support in both wells activating faster transitions. In panel (b),  $T_1$ is plotted versus the flux through the dc-SQUID $\Phi_\mathrm{dc}$ and the effective $E_J$ value for both gate working points and temperatures. The curves display an expected exponential scaling with the potential barrier height, that is proportional to the  value of $E_J$. Notably, the curves for the two flux points show a distinct behaviour. For the working point of the single qubit gate, the scaling is similar, but it depicts a 'staircase' pattern, as observed  experimentally for the double-well system of the Kerr-cat qubit \cite{frattini, kerrcatladder, kerrcat2}. This feature is similarly here attributed to the number of energy eigenstate pairs that are located within the wells. This is clearly absent from the flux point $\Phi_\mathrm{ext}=0.44~ \Phi_0$, for which the potential has a stronger asymmetry and no level-pairing occurs. The additional inset displays curves for more temperatures for the two-qubit working point at $\Phi_\mathrm{ext}=0.44~ \Phi_0$.  
Using the numerically simulated dependence of the fluxon state lifetime $ T_1\propto \text{exp}(\gamma E_J/k_{\mathrm{B}} T_{\mathrm{eff}})$ as a function of $E_J$ and temperature $T_{\mathrm{eff}}$ as shown in the inset of panel (b) we extract a scaling constant $\gamma=1.48 \pm 0.06$, confirming the exponential dependence with $E_J/T_{\mathrm{eff}}$ \cite{fluxcat}.\\

For a numerical estimate of the dephasing time  we use  \cite{Sun2023}

\begin{equation}
T_\phi = (\ln(2) (
\Tilde{A}^2_1 +\Tilde{A}^2_2    + 2 c_{12} \Tilde{A}_1 \Tilde{A}_2))^{-\tfrac{1}{2}}
\end{equation}

with $\Tilde{A}_i=A_i \frac{\partial \omega_{01}}{\partial \Phi_i}$
and the noise correlation coefficient $c_{12}$ is assumed to be $0.5$. The variables refer to the physical fluxes $\Phi_1$ and $\Phi_2$ and are related through  $\Phi_1=\Phi_\mathrm{dc}$ and   $\Phi_2=\Phi_\mathrm{ext}-\Phi_\mathrm{dc}/2$ (Appendix~\ref{app}), with realistic flux noise amplitudes $A_1=A_2= 10 \,\mu\Phi_0$ for regularly-sized SQUIDs \cite{Sun2023}.  \\

The panels (c) and (d) show the calculated values of $T_2=(\tfrac{1}{2T_1}+\tfrac{1}{T_\phi})^{-1}$ versus the  external flux $\Phi_\mathrm{ext}$ and $\Phi_\mathrm{dc}$, respectively. In the heavy regime, $T_2$ is estimated to be around $6.4~\mu s$ for both working points and in the light regime around a factor of 5 higher.  As depicted in (d), $T_2$ is mainly limited by $T_\phi$ in the heavy regime and is fairly constant except for small values of $E_J$. This is attributed to the vanishing double-well structure and therefore the absence of fluxon states approaching small $E_J$ values. 
The panels (e) and (f) in \autoref{coherence_fig} display the energy splitting between the two fluxon states versus $\Phi_\mathrm{ext}$ and $\Phi_\mathrm{dc}$. 

\begin{table}[h]
\caption{Simulated $T_1$ and $T_2$ values at the two gate working-points.}\label{coherence_table}
\begin{tabular}{|lllll|}
\hline
\multicolumn{1}{|l|}{$T_\mathrm{eff}$} & \multicolumn{1}{l|}{$T_1^{\mathrm{heavy}}$ {[}$s${]}} & \multicolumn{1}{l|}{$T_2^{\mathrm{heavy}}$ {[}$\mu s${]}} & \multicolumn{1}{l|}{$T_1^{\mathrm{light}}$ {[}$\mu s${]}} & $T_2^{\mathrm{light}}$ {[}$\mu s${]} \\ \hline
\multicolumn{5}{|c|}{Single-qubit gate ($\Phi_\mathrm{ext} = 0.495~ \Phi_0$)}                                                                            \\ \hline
\multicolumn{1}{|l|}{60 mK} & \multicolumn{1}{l|}{$3.3 $} & \multicolumn{1}{l|}{6.4} & \multicolumn{1}{l|}{$150 $} & 93\\ \hline
\multicolumn{1}{|l|}{70 mK} & \multicolumn{1}{l|}{$0.61$} & \multicolumn{1}{l|}{6.4} & \multicolumn{1}{l|}{$140$} & 87 \\ \hline

\multicolumn{5}{|c|}{Two-qubit gate ($\Phi_\mathrm{ext}= 0.44~ \Phi_0$)}                                                                            \\ \hline
\multicolumn{1}{|l|}{60 mK} & \multicolumn{1}{l|}{$32$} & \multicolumn{1}{l|}{6.4} & \multicolumn{1}{l|}{$150$} & 18 \\ \hline

\multicolumn{1}{|l|}{70 mK} & \multicolumn{1}{l|}{$4.5 $} & \multicolumn{1}{l|}{6.4} & \multicolumn{1}{l|}{$130$} & 18 \\ \hline
\end{tabular}
\end{table}

The comparison of the simulated coherence times for the working points of the single- and two-qubit gate is summarized in \autoref{coherence_table}. $T_1$ in the protected regime is estimated to be in the range of seconds and by a factor $10^5-10^6$ longer as in the light regime. Therefore, on relevant timescales, we expect to be mainly limited by $T_2$ and possible relaxation errors during the short duration of the single and two-qubit gates. Importantly, while an exponential gain in $T_1$ is seen with increasing $E_J$, our simulations indicate that this is not at the cost of reduced $T_2$, in the high $E_J$ region. This is in contrast to the phase coherence scaling in bit-flip protected cat qubits \cite{lescanne, reglade, putterman1, putterman,frattini, siddiqi}.

\section{Conclusion}

Here we proposed a new type of superconducting qubit: a fluxonium qubit with exponentially tunable bit-flip protection. The tunability is based on the flux control of a dc-SQUID that  varies either the value of  $E_J$ or alternatively $E_C$ of the circuit. The qubits are operated in two regimes - the heavy regime for information storage, for which $T_1$ is estimated to show an exponential scaling with the value of $E_J$ and has shown to be even as large as hours \cite{farid_fluxon}. For gate operations, the qubits are brought to the light regime for a nanoscale duration that is small compared to the coherence times. Since the gate implementation is solely based on flux pulses and no microwave lines are needed except for readout, this can be advantageous with regard to scaleability, as frequency crowding will not affect the control capabilities. The main limitation of the proposed qubits, is the short $T_2$ due to a high dephasing rate, which is caused by flux noise. Reducing the flux noise amplitude has been achieved with appropriate surface treatment \cite{surface}  and due to its low frequency nature there are a variety of active mitigation techniques \cite{Gustavsson, Maurer, floquet,thibodeau2024}. Given the strong noise-bias of the proposed qubits, they can be further concatenated with hardware-efficient error-correcting codes that leverage this bias \cite{biased_noise, QEC1, QEC2, QEC3, biased_noise1, putterman}. Long decay times, state-of-the art gate fidelities, and convenient control make the proposed type of qubits a promising approach for a scale-able quantum computing architecture.\\

\textbf{Acknowledgments.} The authors thank Joao Romeiro and  Stefan Filipp for useful discussions, Danyang Chen and Jens Koch for advice to improve the simulation performance. The Python packages SCqubits \cite{scqubits, scqubits2} and Qutip \cite{qutip, qutip2} were used for the numerical simulations.
This work was supported by a NOMIS foundation research grant, the Austrian Science Fund (FWF) through BeyondC F7105,  the Horizon Europe Program HORIZON-CL4-2022-QUANTUM-01-SGA via Project No.~101113946 OpenSuperQPlus100, and IST Austria.

\onecolumngrid

\appendix
\makeatletter
\def\Hy@appendixautorefname{Appendix}
\makeatother

\renewcommand*{\sectionautorefname}{Appendix}

\section{Derivation of the tunable-$E_J$ Hamiltonian} \label{app}

We present the derivation of the time-dependent Hamiltonian for the tunable-$E_J$ fluxonium, in presence of finite squid asymmetry. We start with the Hamiltonian in the irrotational gauge as proposed in \cite{You2023} for correct flux allocation in time-dependant cases

\begin{align}
    \hat{H}_{E_J} =  &E_C (2 \hat{n})^2+\frac{E_L}{2} (\hat{\varphi}-\varphi_{1} \frac{ C_{J_1}}{C_{\Sigma}}-\varphi_{2} \frac{ C_{J_1}+C_{J_2}}{C_{\Sigma}})^2\\\nonumber
    &-E_{J_1}  \cos (\hat{\varphi}+\varphi_{1} \frac{ C_{J_2}+C}{C_{\Sigma}}+\varphi_{2} \frac{ C}{C_{\Sigma}})-E_{J_2}  \cos (\hat{\varphi}-\varphi_{1} \frac{ C_{J_1}}{C_{\Sigma}}+\varphi_{2} \frac{ C}{C_{\Sigma}}),
\end{align}

with  $E_{J_i}$/$C_{J_i}$ the energy/capacitance of the Josephson junction, capacitance $C$ of the inductor and total capacitance of the circuit $C_{\Sigma}=C+C_{J_1}+C_{J_2}$.  $\varphi_{1}$ is the reduced physical flux in the dc-loop and $\varphi_{2}$ in the rf-loop. Given the geometry of the circuit, the effective reduced fluxes of the $\textrm{rf}$-squid and the $\textrm{dc}$-squid are $\varphi_\textrm{ext}=\varphi_{2}+\varphi_{1}/2$ and $\varphi_\textrm{dc}=\varphi_{1}$ and we express the asymmetry of the dc-SQUID as  $E_{J_1/J_2}=E_J(1\pm d)$ and $C_{J_1/J_2}=C_J(1\pm d)$

\begin{align}
    \hat{H}_{E_J} =  &E_C (2 \hat{n})^2+\frac{E_L}{2} (\hat{\varphi}-\varphi_\textrm{dc} \frac{d\, C_j}{C_{\Sigma}}-\varphi_\textrm{ext}\frac{2 C_j}{C_{\Sigma}})^2\\\nonumber
    &-2E_J  \cos (\hat{\varphi}+\varphi_\textrm{ext} \frac{ C}{C_{\Sigma}}-\varphi_\textrm{dc}  \frac{ d\, C_J}{C_{\Sigma}})    \cos (\frac{\varphi_\textrm{dc}}{2}) +2d E_J \sin (\hat{\varphi}+\varphi_\textrm{ext} \frac{ C}{C_{\Sigma}}-\varphi_\textrm{dc}  \frac{ d \,C_J}{C_{\Sigma}})    \sin (\frac{\varphi_\textrm{dc}}{2}).
\end{align}

The latter term due to the finite squid asymmetry $d$ is very harmful to the proposed gate mechanism,  since $\varphi_\textrm{dc}$ is tuned from $0$ to $\pi$. Thereby, the potential of the qubit is changed by quarter of a period over this range, preventing the hybridization of the fluxon states. However, the Hamiltonian can be rewritten \cite{Sun2023} in a more practical form

\begin{equation}
    \hat{H}_{E_J} =E_C (2 \hat{n})^2+\frac{E_L}{2} (\hat{\varphi}-\varphi_\textrm{dc} \frac{d\,C_j}{C_{\Sigma}}-\varphi_\textrm{ext}\frac{2 C_j}{C_{\Sigma}})^2  -E_J(\varphi_\textrm{dc})  \cos ( \hat{\varphi}-\varphi_\textrm{dc} \frac{d\, C_j}{C_{\Sigma}}+\varphi_\textrm{ext}\frac{C}{C_{\Sigma}}+ \varphi_\textrm{corr})  ,
\end{equation}
where we define $E_J(\varphi_\textrm{dc}) = E_{J, \textrm{max}} |\sqrt{\cos^2(\varphi_\textrm{dc}/2) + d^2 \sin^2(\varphi_\textrm{dc}/2)}|$  and $\varphi_\textrm{corr}= \textrm{atan}(d\,  \textrm{tan}(\varphi_\textrm{dc}/2))$.The latter contribution $\varphi_\textrm{corr}$  encapsulates the change of the effective working point for $\varphi_\textrm{dc}$ being modified, 
Thus the dc-SQUID asymmetry $d$ can be mitigated by correcting for this term that changes the effective working point of the fluxonium. 
Since even in absence of a Josephson Junction asymmetry both physical fluxes have to be tuned for independent control of the loops, this adjustment of the correction is straightforward to implement.\\

The Hamiltonian in the main text in \autoref{H_EJn}  underlies two simplifications, that do not alter the  proposed gate mechanism. First, since  $\varphi_\textrm{corr}$ can be corrected for, we absorb it in $\varphi_\textrm{rf}$ and thus simplifying it to the case of $d=0$, but with a finite minimum value of $E_J$. Second, we consider the limit $C_j>>C$ to arrive at the Hamiltonian  in \autoref{H_EJn}
 
\begin{equation}
    \hat{H}_{E_J} =  E_C (2 \hat{n})^2+ \frac{E_L}{2} (\hat{\varphi}-\varphi_\textrm{ext})^2 -E_J(\varphi_\textrm{dc})  \cos ( \hat{\varphi}).
\end{equation}

\twocolumngrid
\bibliography{main}

\end{document}